%% file: lf_candf.tex
\input stbasic.tex

\twelvepoint
\singlespace
\raggedbottom
\null\vskip.5in
\vskip.4in
\centerline{\bf The 
galaxy luminosity function in clusters and the field}
\vskip.4in
\centerline{Neil Trentham }
\smallskip
\medskip
\centerline{Institute for Astronomy, University of Hawaii}
\vskip 4pt
\centerline{2680 Woodlawn Drive, Honolulu HI 96822, U.~S.~A.}
\vskip 4pt
\centerline{email : nat@newton.ifa.hawaii.edu}
\vskip.3in
\vskip.3in
\centerline{Submitted to $MNRAS$} 
\vskip.3in
\vskip.3in
\vfil
\eject
 
\centerline{\bf ABSTRACT }
\bigskip
\noindent
We present the results from a CCD survey of the $B$-band
luminosity functions of
nine clusters of galaxies, and compare them to published photographic
luminosity functions of nearby poor clusters like Virgo and Fornax and
also to the field luminosity function.  We derive a composite luminosity  
function
by taking the weighted mean of all the individual cluster
luminosity functions; 
this composite luminosity function is steep at bright and
faint magnitudes and is shallow in-between.  

All clusters
have luminosity functions consistent with this single composite function.
This is true both for rich clusters like Coma and for poor clusters like
Virgo.

This same composite function is also individually consistent with the
deep field luminosity functions of Cowie et al.~(1996) and 
Ellis et al.~(1996), and also with the 
faint-end of the Las Campanas Redshift Survey $R$-band luminosity
function, shifted by 1.5 magnitudes.
A comparison with the 
Loveday et al.~(1992) field luminosity function which is well-determined  
at the bright-end, shows that   
the composite
function that fits the field data well fainter than $M_B = -19$
drops too steeply between $M_B = -19$ and $M_B = -22$ to fit the field
data there well.  

\bigskip
\noindent{{\bf Key words:} 
galaxies: luminosity function 
$-$ galaxies: clusters: luminosity function} 

\vfil\eject

\noindent{\bf 1 INTRODUCTION}

\noindent
The galaxy luminosity function $\phi(L)$, 
defined as the number density of galaxies
per unit luminosity, is one of the most direct observational probes of
galaxy formation theories.  

Most evidence (see references below) suggests that 
$\phi(L)$ drops steeply brighter than about $M_B = -21$ (i.e., extremely
luminous galaxies are very rare), and rises more gradually fainter than
this.  This form is expressed well by the Schechter (1976) function.
At very faint magnitudes $M_B > -10$, we expect
on theoretical grounds
that $\phi(L)$ turns over again, as 
we expect star formation in very low
mass systems to be suppressed by photoionization of the
intergalactic medium from the ultraviolet
background (Efstathiou 1992, 
Chiba \& Nath 1994, Thoul \& Weinberg 1995), but
this turnover has not yet been observed.

It is also well established that different types of galaxies contribute
proportionately different amounts to $\phi(L)$ at different magnitudes.
At bright magnitudes $M_B > -16$, most galaxies in clusters are giant
ellipticals and in the field they are
giant late-type galaxies.  At fainter
magnitudes $M_B < -16$, most galaxies in clusters are dwarf spheroidal
galaxies 
(see Binggeli 1987 $-$ note that he calls them dwarf ellipticals), 
and in the field, they are either dwarf spheroidal
or dwarf spiral and irregular galaxies.  
The different types of galaxies have different
scaling laws and they lie in different parts of the fundamental
plane parameter 
correlation diagrams (Kormendy 1985, 1987, Binggeli 1994), 
suggesting 
that different physical processes are at work at producing their
luminosities.  There is 
however some overlap between the different types,
particularly between dwarf irregular and dwarf spheroidal 
galaxies.  
That clusters and the field behave differently is thought to
be due to 
the fact that in 
clusters, the crossing time is less than the Hubble time
so that galaxies there
will be influenced by
cluster-related processes.  

What all this suggests is that the physics
controlling the total galaxy luminosity function
is 
very complicated.  Nevertheless $\phi (L)$ 
is a directly observable quantity, so it provides
a popular and direct way to compare theory
with observation.

Early attempts (e.g., White \& Rees 1978) were successful at producing
the general shape of $\phi (L)$.  Recent theoretical advances 
(e.g.~White \& Kauffmann 1994, 
van Kampen 1995, Babul \& Ferguson 1996)
now allow us to calculate 
$\phi(L)$ in detail from first principles albeit with many
assumptions and approximations and thereby
constrain combinations of fundamental
cosmological quantities.  Much observational work in this subject has
been done in both the field (e.g., 
Kirshner et al.~1983, Loveday et al.~1992,
Marzke et al.~1994, Ellis et al.~1996, Cowie et al.~1996) 
and in clusters (e.g., Oemler 1974, Sandage et al.~1985,
Lugger 1986, Oegerle \& Hoessel 1989, Ferguson \& Sandage 1991, 
plus several recent papers $-$ see
Section 2).  This work provides a 
sound basis for comparison with theory. 

Recent advances like the development of
large-format CCDs, 
fibre-optic spectrographs, and the construction of 
the 10 m Keck telescopes, have 
allowed measurement of the galaxy
luminosity down to very faint magnitudes with unprecedented accuracy in
both galaxy clusters and in the field.
In particular, measurements now go deep enough that we can begin to
probe the part of $\phi (L)$ where the dwarf galaxies contribute
significantly.  Here we present a compilation of recent work at these
faint limits and assess the
case for a universal galaxy luminosity function. 

Throughout this paper we assume that
$H_0 = 75$ km s$^{-1}$ Mpc$^{-1}$ and
that $\Omega_0 = 1$.

\vskip 10pt

\vfil\eject

\noindent{\bf 2 OBSERVATIONS AND RESULTS} 

\vskip 5pt

\noindent{\bf 2.1 Clusters}

\noindent
Here we present the results of a study of the (Johnson)
$B$-band luminosity
functions of nine clusters with $0 < z < 0.2$, down to
$M_B = -10$.  This work uses various CCD cameras on
the University of Hawaii
2.2 m Telescope and the Canada-France-Hawaii Telescope; the
data and the details of its reduction and analysis are described
elsewhere (Trentham 1997). 
Various specialized techniques have been used;
these include a detailed characterization of the
background number counts for $B < 26$ (see also Driver et al.~1994a,
1994b, Bernstein et al.~1995), and also techniques for measuring
total magnitudes of galaxies that take into account their detailed
seeing-convolved light distributions (Trentham 1996). 

The clusters observed by us are: Abell 665 ($z = 0.18$),
Abell 963 ($z = 0.21$),
Abell 1146 ($z = 0.14$),
Abell 1795 ($z = 0.06$),
Abell 2199 ($z=0.03$),
Coma, Abell 1367, 
Abell 262, and the NGC 507 Group (all $z=0.02$). 
For these clusters, and also for
Virgo (Sandage et al.~1985), 
Fornax (Ferguson 1989), the Local Group
(van den Bergh 1992), and a number of other nearby groups
(Ferguson \& Sandage 1991, Tully 1988), 
we computed a composite luminosity
function.
In this calculation, we adopt the error estimates
given in the original
sources, we normalize so that the total number of galaxies brighter
than $M_B = -16$ in each cluster is the same, and we sum
the luminosity functions.  In effect, this is a weighted average
of all the individual luminosity functions.
We chose $M_B = -16$ as the normalization threshold because this
is approximately
where the dwarf-giant transition occurs in the fundamental plane 
(Kormendy 1987, Binggeli 1994).  This calculation is therefore not 
sensitive to the dwarf-to-giant ratio or to the
properties of the dwarf population.  

The resulting composite luminosity function is presented in Table 1
and in Figure 1.
The figure shows that $\phi(L)$ is steep at very
bright magnitudes, $M_B < -20$ and also at faint magnitudes, 
$-14 < M_B < -11$.  It is somewhat more shallow in between
but it is still significantly steeper than $\alpha = -1$.
The error bars are small enough that the curvature in this
luminosity function is highly significant, and a power-law does
not provide a good fit to the data over any significant
magnitude range.  
Table 1 also lists the local slope ($\alpha$, where
$\phi (L) \propto L^{\alpha}$ and
$N(M) = - \phi(L) { {{\rm d}L}\over{{\rm d}M}} 
\propto 10^{-0.4 (\alpha+1) M}$) of the luminosity function
described by Figure 1. 
Here $-0.4 (\alpha (M) + 1)$ is the best-fitting 
linear slope at $M$ of 
the three points at $M-1$, $M$, and $M+1$.
It is apparent from the numbers listed in Table 1 that
$\alpha < -1$ everywhere; a flat 
composite luminosity function ($\alpha = -1$) is ruled out at any
magnitude.      
We do not present a luminosity function fainter than $M_B = -11$,
as we have measurements this faint only in 
Abell 1367 and Abell 262.

It is interesting that the luminosity function is continuous
over the magnitude range where there is a break between giants and
dwarfs in the fundamental plane ($M_B \sim -16$).
Indeed, it is flatter in this transition region
than anywhere else.
This is suggestive of a conspiracy in which the giant galaxy
luminosity function is falling as luminosity decreases
by an amount 
almost exactly compensated
for by the rise in the dwarf luminosity function.

Figure 2 shows how the average function presented in Figure 1
compares with the individual luminosity functions of each
cluster.  
That the composite function fits all of the individual data sets
so well is remarkable and provides a strong case for a universal
luminosity function in clusters.
Table 2 provides for each case
the number of degrees of freedom $\nu$ and the
reduced $\chi^2$ 
of the composite luminosity
function that is normalized to provide the best
fit to the data, as well as the probability $P$ of obtaining this
value of $\chi^2$ or lower if this  
luminosity function is the correct representation. 
Little
significance should be attached to low numbers in
the final column because
the true errors are probably more systematic than random.
For example, the Tully and Ferguson-Sandage groups are biassed towards
points near $M_B = -20.5$, as these groups were identified by
searching around such galaxies.  Ignoring the $M_B = -20.5$
point improves the
quality of the fit substantially.
The fit in the Local Group is not good either, but given 
the combination
of selection effects (the
Sagittarius dwarf with $M_B \sim -13$ was discovered only two
years ago; Ibata et al.~1994) 
and poor counting statistics, the quality of the fit 
does not have
much significance.  

Also apparent in comparing Figures 1 and 2 is that, at least for the
more
distant clusters ($z > 0.01$), the luminosity function is poorly
constrained in each cluster (this is in part why the $\chi^2$ values in 
Table 2.2 for these clusters are so low), but in combination the clusters 
constrain the composite function very tightly.   
This is because for each cluster, the uncertainties brought about
by field-to-field variance in the subtracted background are high $-$
it is only when we look at a large sample of these clusters in
combination that the pattern shown in Figure 1 becomes clearly
visible.  Note also that the uncertainty 
$\Delta \alpha$ in deriving $\alpha$ for
each cluster increases rapidly as $\alpha$ increases.  Therefore,
measuring $\alpha$ at the faint end
based on the datasets we present for clusters
like A665 and A963 is extremely difficult: we cannot even 
distinguish
between $\alpha = -1.6$ and $\alpha = -1$ at the faint end
(the value of the composite function here is $\alpha = -1.4$ at
the faintest points).  Only if
$\alpha < -2$ in an
individual cluster could we measure it so better than 0.1. 

It is particularly intriguing that the luminosity functions
of Virgo and
Fornax are so well
fit by the composite function at the faint end.  
The measurements
of $\phi(L)$ in these nearby clusters come from photographic
wide-field surveys (see Table 1 for references). 
The errors here come from
counting statistics, not from a
background subtraction, 
so they are much smaller than those for the more
distant clusters.   In Virgo, the
faintest four points, which use the Impey et al.~(1988)
completeness corrections, are not used in deriving the
composite function.  The agreement between these points and the
composite line is perhaps suggestive that the corrections are
accurate.
The case of Fornax is even more remarkable.  The faintest three
points were not used in computing the composite function, 
because we worried about surface-brightness selection effects
like those described by Impey et al.~(1988) in Virgo.
Yet they agree very well with the composite function.  
Given the very small
errors in the Fornax dataset (Ferguson 1989),
we had no reason to expect $\chi^2$ to be so low; we therefore  
suggest that the surface-brightness selection effects in the
Fornax sample are small at the faint end, 
and, more importantly, that the composite function
is in fact valid over a wide range of richness at faint magnitudes. 
 
Although this paper only presents CCD observations of rich
clusters from our own $B$-band survey
(to ensure consistency in the photometry as well as to
ensure consistency with the photographic and field data 
magnitude systems), a number of other papers have recently
appeared on this subject to which the reader is
referred (Thompson \& Gregory 1993, Driver et al.~1994b,
Biviano et al.~1995, De Propris et al.~1995, 
Bernstein et al.~1995, Secker \& Harris 1996,
Wilson et al.~1997),
Many other authors find results similar to
ours.  
The biggest differences are seen in A2199 (where we find a far shallower
luminosity function than that seen by De Propris et al.~1995; this is
because our field-to-field variance in the background more than an order
of magnitude larger than theirs, so that our uncertainty 
in the faintest points, which define their slope, is much larger) and
A665 (where the large number of faint galaxies predicted by the completeness
model of Wilson et al.~1997 are not seen in our deeper data).
Smaller discrepancies are seen in A963 (where we use a more generalized
method of making isophotal corrections than Driver et al.~1994 which leads
us to measure larger total magnitudes for the faintest galaxies and so derive
a shallower slope) and Coma (where we find more galaxies in our data 
at the faint
end than predicted by the completeness model of Secker \& Harris 1996). 
It is encouraging, however, that
the database of cluster luminosity functions
is growing, particularly at high redshift, where 
the galaxy population in clusters (Butcher \& Oemler
1978, 1984; Dressler et al.~1994)
is quite different to
that at $z=0$. 
 
\vskip 6pt

\noindent{\bf 2.2 Field}

We therefore have a composite luminosity function which seems to
provide very good fits to the data in both rich clusters like Coma
and poor clusters like Virgo and Fornax.
Next we investigate how well it fits the field data, and compare
it to the two deepest datasets (the Autofib dataset of Ellis et al.~1996,
and the Keck dataset of Cowie et al.~1996).  We also compare it to the 
Loveday et al.~(1992) sample, which covers much more area than the
deep samples, but does not sample such faint absolute magnitudes.

Figure 3 shows how our composite cluster function compares to the
field data.  
We find that our function is consistent with
either the Cowie et al.~data or the Autofib data, but with different
normalizations that are
inconsistent with each other at the 3$\sigma$ confidence
level.  
The issue of the normalization of the field luminosity function is
complicated (see e.g.~Colless 1995 and references therein, 
Heyl et al.~1996, Lin et al.~1996)
and there may be many explanations for what is seen here,
for example clustering and large-scale structure along the line of
sight.  One of the classical problems (Colless 1995) of   
the field surveys was their bias towards high surface-brightness  
galaxies; this is unlikely to be a problem for the deep field 
surveys here
as they detect galaxies with lower surface-brightnesses than 
the faintest galaxies in the cluster sample.  

We also compare our luminosity function with the Las Campanas Redshift
Survey luminosity function of Lin et al.~(1996).  
This is the largest field-survey to date, comprising 18678 galaxies,
reaching $M_R = -16$.
We do not present 
in Figure 3 because of the different filter systems (it is in the
$R$-band, whereas our data and the other surveys are in the $B$-band).  
For the Lin et al.~sample, in the range $-18 < M_R < -16$ (approximately
$-16.5 < M_B < -14.5$), a power-law fit to their luminosity function
gives $\alpha = -1.39 \pm 0.11 $, which is consistent with the 
value $\alpha = -1.43 $ in this magnitude range for the cluster
composite function (see Table 1).   
However, there is some evidence
the Lin et al.~luminosity function flattens at their very faint-end:
for $-17.5 < M_R < -16$, a power-law fit gives $\alpha = -1.19 \pm 0.19$. 
If this trend continues to fainter magnitudes $-16 > M_R > -12$, 
the field and cluster
luminosity functions would then be incompatible.  There is weak evidence
for this from the Local Group luminosity function, but this might
simply be due to completeness effects. 

In summary, at the faint-end, the composite cluster luminosity
function has a shape consistent with those of the Cowie et al.,
Ellis et al., and Lin et al.~field luminosity functions.  However,
the statistics in the field samples are at present poor, so that
the case for a universal luminosity function at faint magnitudes
cannot be addressed rigorously at present. 

In comparison with the 
Loveday et al.~(1992) field luminosity function which has excellent
statistics at the bright-end, 
the composite
function that fits the data well fainter than $M_B = -19$
drops too steeply between $M_B = -19$ and $M_B = -22$ to fit the field
data there well. 
This could be because massive late-type galaxies whose
dominated by  
young stellar populations, exist in the field but not in clusters
because cluster-related process like 
gas stripping have turned off star formation there. 
We see this effect in the $B$-band because the $B$-band light of
a galaxy primarily comes from the young stars. 
There may also be a deficiency of star-forming galaxies in clusters
at fainter magnitudes, but this does not cause a noticeable difference
in the luminosity
function, because it is much flatter there, and because the
field statistics are worse. 
We also note that the bright-end slope of 
the rich-cluster luminosity function is slightly steeper than that of
the poor-cluster luminosity function (Figure 4), which could also
be due to the same effect. 
 
\vskip 10pt
\noindent{\bf ACKNOWLEDGMENTS}

\noindent
Helpful discussions with Len Cowie and John Kormendy
are gratefully acknowledged.

I also thank Richard Ellis and the staff of the Insitute of Astronomy
in Cambridge for their hospitality during 1996, when most of this work
was carried out.

\noindent
This research has made use of the NASA/IPAC extragalactic database (NED) which
is operated by the Jet Propulsion Laboratory, Caltech, under agreement with the
National Aeronautics and Space Administration.

\vskip 20pt

\ni{\bf REFERENCES }
\beginrefs
 
Abell G.~O., 1958, ApJS, 3, 211
  
Babul A., Ferguson H.~C., 1996, ApJ, 458 100

Bernstein G.~M., Nichol R.~C., Tyson J.~A., Ulmer M.~P., Wittman D., 1995, AJ,
110, 1507

Binggeli B., 1987, in Faber S.~M., ed., Nearly Normal Galaxies.
Springer-Verlag, New York, p.~195

Binggeli B., 1994, in Meylan G., Prugneil P., ed., ESO
Conference and Workshop Proceedings No.~49: Dwarf Galaxies. 
European Space Observatory, Munich, p.~13
 
Binggeli B., Sandage A., Tamman G.~A., 1988, ARAA, 26, 509

Biviano A., Durret F., Gerbal D., Le Fevre O., Lobo C., Mazure A., 
Slezak E., 1995, A\&A, 297, 610 

Butcher H.~R., Oemler, A., 1978, ApJ, 219, 18
  
Butcher H., Oemler A., 1984, ApJ, 285, 426  

Chiba M., Nath B.~B., 1994, ApJ, 436, 618

Coleman G.~D., Wu C-C., Weedman D.~W., 1980, ApJS, 43, 393

Colless M., 1995 in Maddox S.~J., Arag{\'o}n-Salamanca., ed., 
Wide-Field Spectroscopy and the Distant Universe, proceedings of the
35th Herstmonceux Conferencee, World Scientific, Singapore,
p.~263

Cowie L.~L., Songaila A, Hu E.~M., Cohen J.~G., 1996, AJ, 112, 839 

De Propris R., Pritchet C.~J., Harris W.~E., McClure R.~D., 1995,
ApJ, 450, 534

Dressler, A., Oemler A., Sparks W.~B., Lucas R.~A., 1994, ApJ, 435, L23

Driver S.~P., Phillipps S., Davies J.~I., Morgan I.,
Disney M.~J, 1994a, MNRAS, 266, 155 

Driver S.~P., Phillipps S., Davies J.~I., Morgan I.,
Disney M.~J, 1994b, MNRAS, 268, 393  
  
Efstathiou G., 1992, MNRAS, 256, 43p

Ellis R.~S., Colless M., Broadhurst T., Heyl J., Glazebrook K.,
1996, MNRAS, 280, 235 

Ferguson H.~C., 1989, AJ, 98, 367 

Ferguson H.~C., Sandage A., 1991, AJ, 101, 765
  
Heyl J., Colless M., Ellis R.~S., Broadhurst T., 1996, MNRAS, in press

Hodge P.~W., 1989, ARAA, 27, 139 

Ibata R.~A., Gilmore G., Irwin M.~J., 1994, Nat, 370, 194

Impey C., Bothun G., Malin D., 1988, ApJ, 330, 634
  
Kirshner R.~P., Oemler A., Schechter P.~L., Shectman S.~A., 1983, 
AJ, 88, 1285

Kormendy J., 1985, ApJ, 295, 73

Kormendy J., 1987, in Faber S.~M. ed., Nearly Normal Galaxies. 
Springer-Verlag, New York, p.~163
 
Lin H., Kirshner R.~P., Shectman S.~A., Landy S.~D., Oemler A.,
Tucker D.~L., Schechter P.~L., 1996, ApJ, 464, 60

Loveday J., Peterson B.~A., Efstathiou G., Maddox S., 1992, ApJ, 390, 338

Lugger P., 1986, ApJ, 303, 535
 
Maddox S.~J., Efstathiou G., Sutherland W.~J., 1990, MNRAS, 246, 433
 
Marzke R.~O, Geller M.~J., Huchra J.~P., Corwin H.~G., 1994,
AJ, 108, 437

Metcalfe N., Fong R., Shanks T., 1995, MNRAS, 274, 769 
 
Oegerle W.~R., Hoessel J.~G., 1989, AJ, 98, 1523
  
Oemler A., 1974, ApJ, 194, 1

Sandage A., Binggeli B., Tammann G.~A., 1985, AJ, 90, 1759

Schechter P., 1976, ApJ, 203, 297

Secker J., Harris W.~E., 1996, ApJ, 469, 623 

Thompson L.~A., \& Gregory S.~A., 1993, AJ, 106, 2197

Thoul A.~A., Weinberg D.~H., 1995, ApJ, 442, 480

Trentham N., 1996, MNRAS, in press

Trentham N., 1997, PhD Thesis, University of Hawaii

Tully R.~B., 1988, AJ, 96, 73 
  
van den Bergh S., 1992, A\&A, 264, 75 

van Kampen E., 1995, MNRAS, 273, 295 

White S.~D.~M, Kauffmann G., in Munoz-Tunon C., Sanchez F., eds.,
The Formation and Evolution of Galaxies.~Cambridge University Press,
Cambridge, p.~455
   
White S.~D.~M., Rees M.~J., 1978, MNRAS, 183, 321 

Wilson G., Smail I., Ellis R.~S., Couch W.~J., 1997, MNRAS, in press

\endrefs

\vskip 20pt
 
\ni {\bf FIGURE CAPTIONS}
\vskip 10pt
\ni {\bf Figure 1.~}The composite cluster
luminosity function computed as described
in the text and in Table 1. 
Here $N_{\rm gal} \propto \phi (M) = \phi(L)
{{ {\rm d}L}\over{{\rm d}M} }$ 
is the number of galaxies per unit area per unit
magnitude; it is arbitrarily  
normalized to represent the number of galaxies in
a typical Abell (1958) richness 2 cluster.

\vskip 10pt
\ni {\bf Figure 2.~} 
Comparison between each cluster luminosity function and the
composite function.  The lines are the function in
Table 1 and Figure 1, normalized for each cluster to minimize
the scatter with the data for $M_B > -22$.  
In each diagram the data comes from the
source listed in Table 1.  In each case the errors bars are a 
combination of the uncertainties from counting statistics,
field-to-field variance in the background (if a background
subtraction is made), and uncertainties in the galaxy
magnitudes; the reader is referred to the original sources for
details.   The two lines near the top
center in each figure 
represent the slopes $\alpha = -1$ (solid line) and
$\alpha = -2$ (dotted line).  
In the Virgo plot, the open circles represent data faintward
of the completeness limit of Sandage et al.~(1985).
These are data points 
computed by Impey et al.~(1988) using the Sandage data set with
completeness corrections appropriate 
to their $\mu_{\rm lim} = 25.8$ model.   
For the Ferguson-Sandage groups and the Tully groups, the
solid lines represent the composite function normalized to all
the data, and the dashed lines represent the function
normalized to the data minus to $M_B = -20.4$ (Ferguson-Sandage)
or $M_B = -20.5$ (Tully) points.  In the Local Group, only data
brightward of the Sagittarius dwarf completeness limit 
(marked S in the diagram) are 
included in computing the normalization.  

\vskip 10pt
\ni {\bf Figure 3.~}The field luminosity functions of
Cowie et al.~(1996), Ellis et al.~(1996 $-$ the Autofib
survey), and Loveday et al.~(1992).  The composite
luminosity functions normalized to minimize the scatter
with the Cowie, Autofib, and Loveday
(for $M_B > -21$ only) data are also
presented.  Symbols and line styles are as given in
the figure legend.   

The $b_J$ data of Ellis et al.~and
Loveday et al.~have been converted to the $B$ band
using the equation $b_J = B - 0.28 (B-V)$ (Maddox
et al.~1990, see also Metcalfe et al.~1995),
the local galaxy type distribution as a function of
absolute magnitude in the field
of Binggeli et al.~(1988), and
the zero redshift colours of different galaxy
types of Coleman et al.~(1980).
These corrections are typically 0.1 $-$ 0.3 
magnitudes.

\vskip 10pt
\ni {\bf Figure 4.~}The slope of the luminosity function
at $M_B = -20$, computed as described in the text, as
a function of the local galaxy density.
The absolute values of the density are somewhat arbitrary
as different angular regions of different clusters are
surveyed.  Nevertheless we do see a positive correlation
at a weak level as described in the text. 

For the rich clusters, a weighted average of the
slopes and densities for Coma, Abell 1146, Abell 963, and
Abell 665 is used.  For the poor clusters, a weighted 
average of the slopes and densities for Virgo and Fornax
is used.  The solid line represents the slope of the
Loveday et al.~field luminosity function, computed the
same way, and the dotted lines its 1$\sigma$ uncertainty.

\par\vfill\eject\bye

%% file: stbasic.tex
\message{STBASIC.TEX TeX Macro Library}
\message{ }





\def\beginrefs{\begingroup\parindent=0pt\frenchspacing
   \parskip=1pt plus 1pt minus 1pt\interlinepenalty=1000\pretolerance=10000
   \hyphenpenalty=10000\everypar={\hangindent=0.42in}       
  \def\aa##1{{\it Astr.~Ap., \bf ##1}}
  \def\aasup##1{{\it Astr.~Ap.~Suppl., \bf ##1}}
  \def\aasupp##1{{\it Astr.~Ap.~Suppl., \bf ##1}}
  \def\aj##1{{\it A.~J., \bf ##1}}
  \def\annrev##1{{\it Ann.~Rev.\ Astr.~Ap., \bf ##1}}     
  \def\araa##1{{\it Ann.~Rev.\ Astr.~Ap., \bf ##1}}     
  \def\apj##1{{\it Ap.~J., \bf ##1}}     
  \def\apjl##1{{\it Ap.~J. (Letters), \bf ##1}}
  \def\apjlett##1{{\it Ap.~J. (Letters), \bf ##1}}
  \def\apjlet##1{{\it Ap.~J. (Letters), \bf ##1}}
  \def\apjsup##1{{\it Ap.~J.~Suppl., \bf ##1}}
  \def\apjsupp##1{{\it Ap.~J.~Suppl., \bf ##1}}
  \def\baas##1{{\it Bull.~A.A.S., \bf ##1}}
  \def\ban##1{{\it B.A.N., \bf ##1}}
  \def\ibvs##1{{\it Inf. Bull. Var. Stars}, No.~##1}
  \def\mn##1{{\it M.N.R.A.S., \bf ##1}}
  \def\mnras##1{{\it M.N.R.A.S., \bf ##1}}
  \def\pasp##1{{\it Pub.~A.S.P., \bf ##1}}
  \def\ajpasp##1{{\it Pub.~A.S.P., \bf ##1}}
  \def\nat##1{{\it Nature, \bf ##1}}
  \def\nature##1{{\it Nature, \bf ##1}}}

\def\endrefs{\endgroup}



\def\df{\leaders\hbox to 0.6em{\hss.}\hfill}


\def\section#1{\bigbreak\medskip\centerline{#1}\par\nobreak\medskip\markpage}

\def\subsection#1#2{\bigbreak\noindent{\bf#1\hskip 0.9em\relax#2}\par
   \nobreak\medskip\markpage}

\def\subsubsection#1#2{\medbreak\noindent{\sl#1\hskip 0.60em\relax#2}\par
   \nobreak\medskip\markpage}

\def\today{\advance\year by -1900 
   \number\month/\number\day/\number\year}
\def\yearmonthday{\number\year\space
   \ifcase\month\or January\or February\or March\or April\or May\or June\or
   July\or August\or September\or October\or November\or December\fi
   \space\number\day}

\newcount\num

\def\nextnum{\global\advance \num by 1 \number\num}
\def\nextitem{\leavevmode
   \hbox{\ifnum\num>8 \kern-0.43em\fi \nextnum.\kern0.60em}}
\def\bfnextitem{\leavevmode
   \hbox{\ifnum\num>8 \kern-0.43em\fi \bf\nextnum.\kern0.60em}}

\newcount\colnum

\def\nextcolnum{\global\advance \colnum by 1 \number\colnum}
\def\nextcolumn{\leavevmode
   \hbox{{\it \ifnum\colnum<9 \phantom{1}\fi Column \nextcolnum:}\kern0.60em}}

\newcount\fig

\def\nextfig{\global\advance \fig by 1 \number\fig}

\newcount\cap

\def\nextcap{\global\advance \cap by 1 \number\cap}

\newcount\letter

\def\nextlet{\global\advance \letter by 1
   \ifcase\letter\or A\or B\or C\or D\or E\or F\or G\or H\or I\or
   J\or K\or L\or M\or N\or O\or P\or Q\or R\or S\or T\or U\or V\or W\or X\or
   Y\or Z\fi}

\newdimen\bigindent \bigindent=3.5in
\def\letterhead{\hsize=6in\interlinepenalty=2000\parskip=6pt minus 3pt
  \pretolerance=750
  \def\topline##1{\hbox to\hsize{\hfil##1\hskip\rightskip}}
  \footline={\ifnum\pageno=1
    \hss\hbox{\vrule height 0.4in width 0pt}
    \eightrm Operated by the Association of Universities for Research in 
    Astronomy, Inc., for the National Aeronautics and Space Administration\hss
    \else\hfil\fi}
  \null
  \vskip-0.2in
  {\advance\rightskip by -0.75in
    \topline{3700 San Martin Drive}
    \topline{Baltimore, MD 21218}
    \topline{(301) 338-4718}\par}
  \vskip30pt minus 15pt
  {\leftskip=\bigindent\yearmonthday\par}}

\def\arpanetletterhead{\hsize=6in\interlinepenalty=2000\parskip=6pt minus 3pt
  \pretolerance=750
  \def\topline##1{\hbox to\hsize{\hfil##1\hskip\rightskip}}
  \footline={\ifnum\pageno=1
    \hss\hbox{\vrule height 0.4in width 0pt}
    \eightrm Operated by the Association of Universities for Research in 
    Astronomy, Inc., for the National Aeronautics and Space Administration\hss
    \else\hfil\fi}
  \null
  \vskip-0.2in\vskip-3\baselineskip
  {\advance\rightskip by -0.75in
    \topline{3700 San Martin Drive}
    \topline{Baltimore, MD 21218}
    \topline{(301) 338-4718}
    \topline{{\elevenrm BITNET:} \tt golombek@stsci}
    \topline{\elevenrm SPAN: \tt SCIVAX::GOLOMBEK}
    \topline{{\elevenrm ARPANET:} \tt golombek@scivax.arpa}\par}
  \vskip30pt minus 15pt
  {\leftskip=\bigindent\yearmonthday\par}}

\def\gosbletterhead{\hsize=6in\interlinepenalty=2000\parskip=6pt minus 3pt
  \pretolerance=750
  \def\topline##1{\hbox to\hsize{\hfil##1\hskip\rightskip}}
  \footline={\ifnum\pageno=1
    \hss\hbox{\vrule height 0.4in width 0pt}
    \eightrm Operated by the Association of Universities for Research in 
    Astronomy, Inc., for the National Aeronautics and Space Administration\hss
    \else\hfil\fi}
  \null
  \vskip-0.375in
  {\advance\rightskip by -0.75in
    \topline{General Observer Support Branch}
    \topline{3700 San Martin Drive}
    \topline{Baltimore, MD 21218}
    \topline{(301) 338-4996}\par}
  \vskip30pt minus 15pt
  {\leftskip=\bigindent\yearmonthday\par}}



\def\indentleft{\advance\leftskip by 50pt\interlinepenalty=750}
\def\inndentleft{\advance\leftskip by 78pt\interlinepenalty=750}
\def\narrower{\advance\leftskip by 0.42in\advance\rightskip by 0.42in
  \interlinepenalty=750}
\def\nnarrower{\advance\leftskip by 50pt\advance\rightskip by 45pt
  \interlinepenalty=750}

\def\checkbox{\nnarrower\parindent=0pt\itemitem{\vbox{\hrule height.7pt
  \hbox{\vrule width.7pt height6pt \kern6pt \vrule width.7pt}
  \hrule height.7pt}$\,$}}  


%
%
\newcount\index \index=100
\def\markpage{\advance\index by 1 \count\index=\pageno}
\def\begintableofcontents{\begingroup
  \index=100 \frenchspacing\interlinepenalty=750
  \parskip=0.1pt plus 1pt minus 0.1pt \parindent=0.3in
  \def\dfi{\advance\index by 1 \df\number\count\index}
  \def\in{\par\hskip-0.2in\indent \hangindent2\parindent \textindent}    
  \def\inin{\par\hskip0.32in\indent \hangindent3\parindent \textindent}
  \def\ininin{\par\hskip0.95in\indent \hangindent4\parindent \textindent}}



{\obeylines\gdef\startdisplay#1
  {\catcode`\^^M=5$$#1\halign\bgroup\indent##\hfil&&\qquad##\hfil\cr}}
\outer\def\enddisplay{\crcr\egroup$$}

\chardef\other=12

{\obeyspaces\gdef {\ }} 

  \font\twentyfourrm=cmr10 scaled 2488
  \font\twentyfouri=cmmi10 scaled 2074   
  \font\twentyfoursy=cmsy10 scaled 2074
  \font\twentyrm=cmr10 scaled 2074      
  \font\twentyi=cmmi10 scaled 2074   
  \font\twentysy=cmsy10 scaled 2074
  \font\eighteenrm=cmr10 scaled 1728
  \font\eighteeni=cmmi10 scaled 1728 \font\eighteensy=cmsy10 scaled 1728
  \font\fourteenrm=cmr10 scaled 1440
  \font\fourteeni=cmmi10 scaled 1440 \font\fourteensy=cmsy10 scaled 1440
  \font\twelverm=cmr12
                
  \font\twelvei=cmmi12               \font\twelvesy=cmsy10 scaled 1200
  \font\elevenrm=cmr10 scaled 1095
    
  \font\eleveni=cmmi10 scaled 1095   \font\elevensy=cmsy10 scaled 1095
  \font\tenrm=cmr10
                   
  \font\teni=cmmi10  \font\tensy=cmsy10  
  \font\ninerm=cmr9

  \font\ninei=cmmi9                  \font\ninesy=cmsy9
  \font\eightrm=cmr8
  \font\seveni=cmmi7 \font\sevensy=cmsy7

\def\commonstuff{
  \parindent=0.42in       
  \def\skipline{\vskip\baselineskip}
  \hyphenpenalty=200\pretolerance=300\tolerance=600 
  \interlinepenalty=100\clubpenalty=500\widowpenalty=500
  \nonfrenchspacing\singlespace\rm}

\def\twelvepoint{
  \font\bf=cmbx12
  \font\it=cmti12
  \font\sl=cmsl12
  \font\tb=cmtt10 scaled 1200 
  \font\tt=cmtt8 scaled 1440
  \textfont0=\twelverm \scriptfont0=\tenrm     
    \scriptscriptfont0=\sevenrm                 
  \def\rm{\fam0 \twelverm}   
  \textfont1=\twelvei  \scriptfont1=\teni  
    \scriptscriptfont1=\seveni                  
  \def\mit{\fam1 } \def\oldstyle{\fam1 \twelvei}
  \textfont2=\twelvesy \scriptfont2=\tensy 
    \scriptscriptfont2=\sevensy                 
  \def\singlespace{\baselineskip=13.5pt\lineskiplimit=-5pt
    \lineskip=0pt
    \parskip=1.25pt plus 1.5pt minus 0.25pt}  
  \def\oneandahalfspace{\baselineskip=18pt\parskip=0pt plus 1pt}
  \def\doublespace{\baselineskip=24pt\parskip=0pt plus 0.5pt}
  \footline={\ifnum\pageno=1 \hfil
             \else\hss\twelverm-- \folio\ --\hss\fi} 
  \def\pagenumbers{\footline={\hss\twelverm-- \folio\ --\hss}}  
  \def\romanpagenumbers{\footline={\hss\twelverm-- \romannumeral\folio\ --\hss}}
  \commonstuff}

\def\tenpoint{
  \font\it=cmti10
  \font\sl=cmsl10
  \font\bf=cmb10
  \textfont0=\tenrm \scriptfont0=\sevenrm     
    \scriptscriptfont0=\fiverm                 
  \def\rm{\fam0 \tenrm}   
  \textfont1=\teni  \scriptfont1=\seveni  
    \scriptscriptfont1=\fivei                  
  \def\mit{\fam1 } \def\oldstyle{\fam1 \teni}
  \textfont2=\tensy \scriptfont2=\sevensy 
    \scriptscriptfont2=\fivesy                 
  \def\singlespace{\baselineskip=12pt\lineskiplimit=0pt
    \lineskip=-0.5mm       
    \parskip=2pt plus 1pt minus 1pt}  
  \footline={\ifnum\pageno=1 \hfil
             \else\hss\tenrm-- \folio\ --\hss\fi} 
  \def\oneandahalfspace{\baselineskip=18pt\parskip=0pt plus 1pt}
  \def\doublespace{\baselineskip=24pt\parskip=0pt plus 1 pt}
  \def\pagenumbers{\footline={\hss\tenrm-- \folio\ --\hss}}  
  \def\romanpagenumbers{\footline={\hss\tenrm-- \romannumeral\folio\ --\hss}}
  \commonstuff}

\def\elevenpoint{
  \font\it=cmti10 scaled 1095
  \font\sl=cmsl10 scaled 1095
  \font\bf=cmb10 scaled 1095 
  \font\tt=cmtt10 scaled 1095
  \textfont0=\elevenrm \scriptfont0=\tenrm     
    \scriptscriptfont0=\ninerm                 
  \def\rm{\fam0 \elevenrm}   
  \textfont1=\eleveni  \scriptfont1=\teni  
    \scriptscriptfont1=\ninei                  
  \def\mit{\fam1 } \def\oldstyle{\fam1 \eleveni}
  \textfont2=\elevensy \scriptfont2=\tensy 
    \scriptscriptfont2=\ninesy                 
  \def\singlespace{\baselineskip=13pt\lineskiplimit=-5pt
    \lineskip=0mm       
    \parskip=2pt plus 1pt minus 1pt}  
  \footline={\ifnum\pageno=1 \hfil
             \else\hss\elevenrm-- \folio\ --\hss\fi} 
  \def\oneandahalfspace{\baselineskip=19pt\parskip=0pt plus 1pt}
  \def\doublespace{\baselineskip=26pt\parskip=0pt plus 1 pt}
  \def\pagenumbers{\footline={\hss\elevenrm-- \folio\ --\hss}}  
  \def\romanpagenumbers{\footline={\hss\tenrm-- \romannumeral\folio\ --\hss}}
  \commonstuff}

\def\eighteenpoint{           
  \font\bf=cmbx10 scaled 1728
  \font\it=cmti10 scaled 1728
  \font\sl=cmsl10 scaled 1728
  \font\tb=cmtt10 scaled 1728
  \font\tt=cmtt10 scaled 1728
  \textfont0=\eighteenrm \scriptfont0=\fourteenrm
    \scriptscriptfont0=\twelverm                 
  \def\rm{\fam0 \eighteenrm}   
  \textfont1=\eighteeni  \scriptfont1=\fourteeni  
    \scriptscriptfont1=\twelvei                  
  \def\mit{\fam1 } \def\oldstyle{\fam1 \eighteeni}
  \textfont2=\eighteensy \scriptfont2=\fourteensy 
    \scriptscriptfont2=\twelvesy                 
  \def\singlespace{\baselineskip=21pt\lineskiplimit=-5pt
    \lineskip=0pt
    \parskip=4pt plus 1pt minus 1pt}  
  \def\oneandahalfspace{\baselineskip=30pt\parskip=0pt plus 1pt}
  \def\doublespace{\baselineskip=40pt\parskip=0pt plus 1pt}
  \footline={\ifnum\pageno=1 \hfil
             \else\hss\eighteenrm-- \folio\ --\hss\fi} 
  \def\pagenumbers{\footline={\hss\eighteenrm-- \folio\ --\hss}}  
  \commonstuff}

\def\twentypoint{
  \font\bf=cmbx10 scaled 2074
  \font\it=cmti10 scaled 2074
  \font\sl=cmsl10 scaled 2074
  \font\tb=cmtt10 scaled 2074
  \font\tt=cmtt10 scaled 2074
  \textfont0=\twentyrm \scriptfont0=\eighteenrm     
    \scriptscriptfont0=\fourteenrm                 
  \def\rm{\fam0 \twentyrm}   
  \textfont1=\twentyi  \scriptfont1=\eighteeni  
    \scriptscriptfont1=\fourteeni                  
  \def\mit{\fam1 } \def\oldstyle{\fam1 \twentyi}
  \textfont2=\twentysy \scriptfont2=\eighteensy 
    \scriptscriptfont2=\fourteensy                 
  \def\singlespace{\baselineskip=24pt\lineskiplimit=-5pt
    \lineskip=0pt
    \parskip=5pt plus 1.5pt minus 1.5pt}  
  \def\oneandahalfspace{\baselineskip=33pt\parskip=0pt plus 1pt}
  \def\doublespace{\baselineskip=44pt\parskip=0pt plus 0.5pt}
  \footline={\ifnum\pageno=1 \hfil
             \else\hss\twentyrm-- \folio\ --\hss\fi} 
  \def\pagenumbers{\footline={\hss\twentyrm-- \folio\ --\hss}}  
  \def\romanpagenumbers{\footline={\hss\twentyrm-- \romannumeral\folio\ --\hss}}
  \commonstuff}

\def\twentyfourpoint{
  \font\bf=cmbx10 scaled 2488
  \font\it=cmti10 scaled 2488
  \font\sl=cmsl10 scaled 2488
  \font\tb=cmtt10 scaled 2488
  \font\tt=cmtt10 scaled 2488
  \textfont0=\twentyfourrm \scriptfont0=\twentyrm     
    \scriptscriptfont0=\eighteenrm                 
  \def\rm{\fam0 \twentyfourrm}   
  \textfont1=\twentyfouri  \scriptfont1=\twentyi  
    \scriptscriptfont1=\eighteeni                  
  \def\mit{\fam1 } \def\oldstyle{\fam1 \twentyfouri}
  \textfont2=\twentyfoursy \scriptfont2=\twentysy 
    \scriptscriptfont2=\eighteensy                 
  \def\singlespace{\baselineskip=28pt\lineskiplimit=-5pt
    \lineskip=0pt
    \parskip=5pt plus 1.5pt minus 1.5pt}  
  \def\oneandahalfspace{\baselineskip=42pt\parskip=0pt plus 1pt}
  \def\doublespace{\baselineskip=56pt\parskip=0pt plus 0.5pt}
  \footline={\ifnum\pageno=1 \hfil
             \else\hss\twentyfourrm-- \folio\ --\hss\fi} 
  \def\pagenumbers{\footline={\hss\twentyfourrm-- \folio\ --\hss}}  
  \def\romanpagenumbers{\footline={\hss\twentyfourrm-- \romannumeral\folio\ --\hss}}
  \commonstuff}

\def\spose#1{\hbox to 0pt{#1\hss}}
\def\lta{\mathrel{\spose{\lower 3pt\hbox{$\mathchar"218$}}
     \raise 2.0pt\hbox{$\mathchar"13C$}}}
\def\gta{\mathrel{\spose{\lower 3pt\hbox{$\mathchar"218$}}
     \raise 2.0pt\hbox{$\mathchar"13E$}}}

\def\ni{\noindent}
\def\in{\indent}
\def\inin{\in{\in}
\def\ininin{\inin{\in}}}
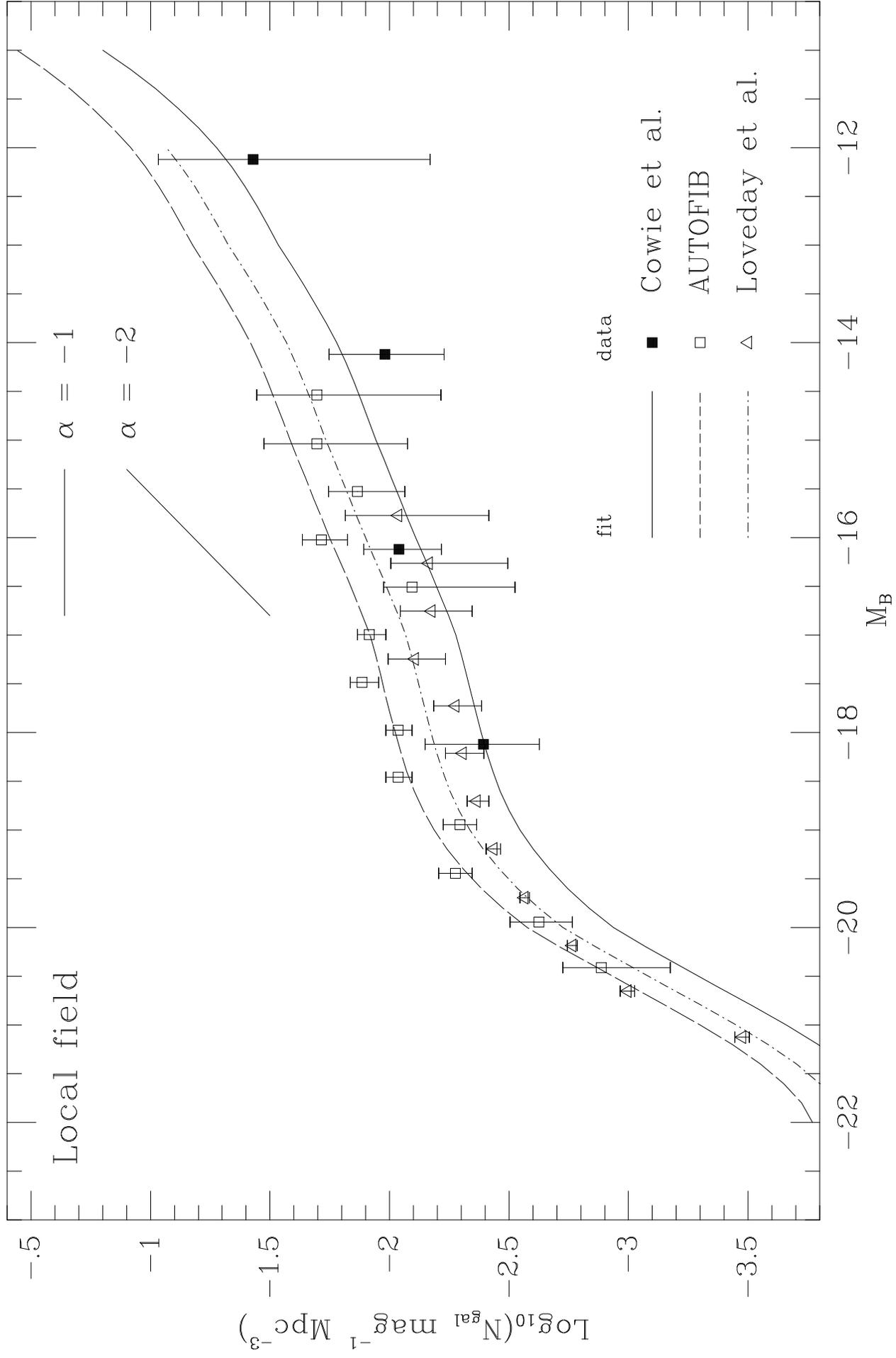
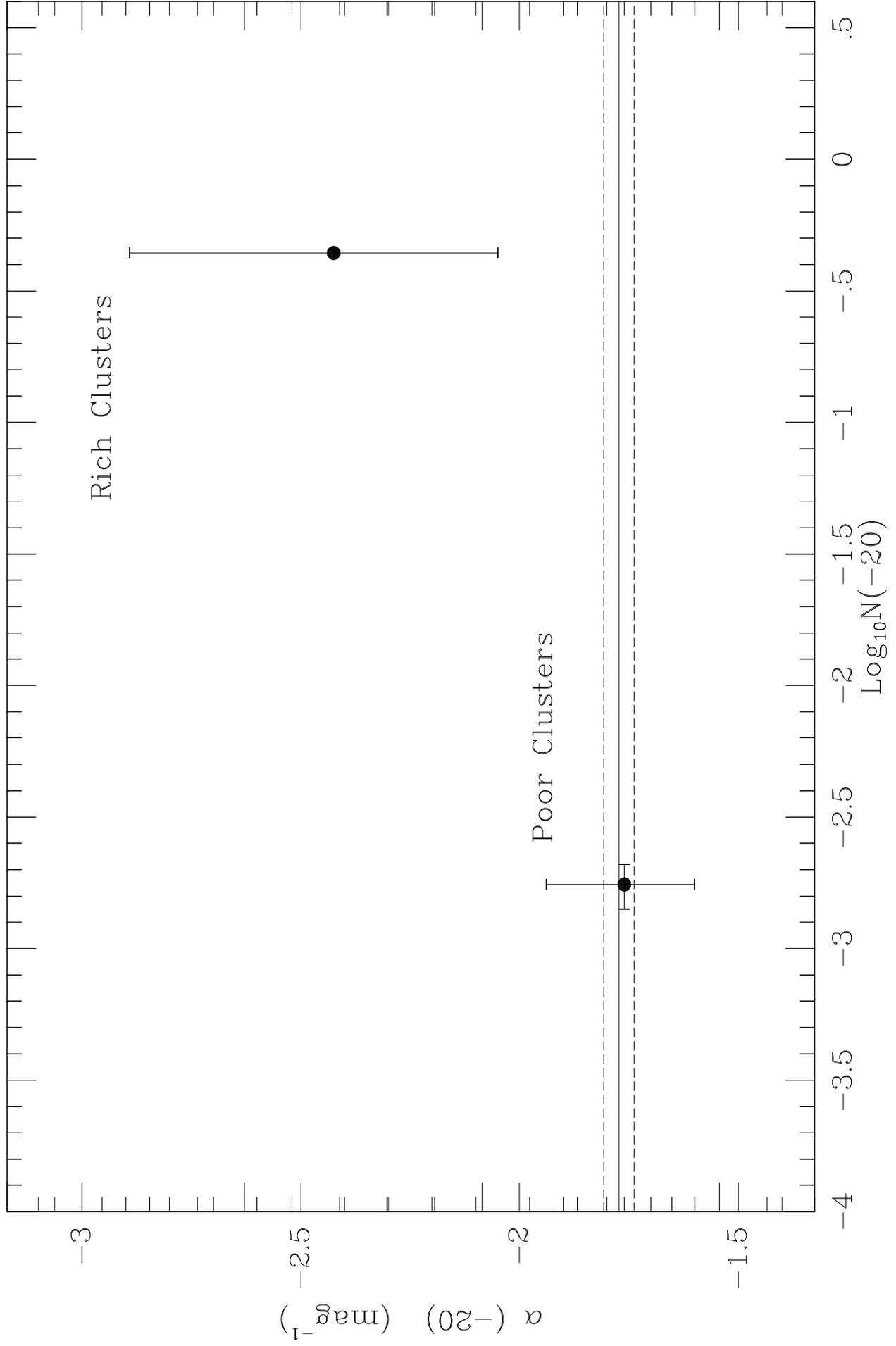
